\documentclass[sigconf]{acmart}

\usepackage{booktabs} 
\usepackage{graphicx}  
\usepackage{amsmath}
\usepackage{multicol}
\usepackage[labelfont=bf]{caption}
\usepackage{subfig}
\usepackage{float}
\usepackage{amsfonts,algorithmic,algorithm}
\usepackage{booktabs,color}
\usepackage{multirow}
\usepackage{enumitem}
\usepackage{caption}
\usepackage{graphicx,booktabs}
\usepackage{graphicx}
\usepackage{caption}
\usepackage{mathtools}

\newtheorem{theorem}{Problem}

\setcopyright{rightsretained}
\usepackage{amsmath}
\usepackage{multicol}
\usepackage[labelfont=bf]{caption}
\usepackage{subfig}
\usepackage{float}
\usepackage{amsfonts,algorithmic,algorithm}
\usepackage{booktabs,color}
\usepackage{multirow}
\usepackage{enumitem}

\copyrightyear{2020}
\acmYear{2020}
\setcopyright{acmcopyright}
\acmConference[WSDM '20]{The Thirteenth ACM International Conference on Web Search and Data Mining}{February 3--7, 2020}{Houston, TX, USA}
\acmBooktitle{The Thirteenth ACM International Conference on Web Search and Data Mining (WSDM '20), February 3--7, 2020, Houston, TX, USA}
\acmPrice{15.00}

\begin{document}
\title[Privacy-Aware Recommendation with Private-Attribute Protection]{Privacy-Aware Recommendation with Private-Attribute Protection using Adversarial Learning}

\author{Ghazaleh Beigi,  Ahmadreza Mosallanezhad, Ruocheng Guo, Hamidreza Alvari, Alexander Nou, Huan Liu}
\affiliation{%
	\institution{Computer Science and Engineering, Arizona State University, Tempe, Arizona}
}
\email{{gbeigi, amosalla, rguo12, halvari, asnou, huan.liu}@asu.edu}

\renewcommand{\shortauthors}{G. Beigi et al.}

\begin{abstract}
Recommendation is one of the critical applications that helps users find information relevant to their interests. However, a malicious attacker can infer users' private information via recommendations. Prior work obfuscates user-item data before sharing it with recommendation system. This approach does not explicitly address the quality of recommendation while performing  data obfuscation. Moreover, it cannot protect users against private-attribute inference attacks based on recommendations. This work is the first attempt to build a \textit{R}ecommendation with \textit{A}ttribute \textit{P}rotection (\textsc{RAP}) model which simultaneously recommends relevant items and counters private-attribute inference attacks. The key idea of our approach is to formulate this problem as an adversarial learning problem with two main components: the private attribute inference attacker, and the Bayesian personalized recommender. The attacker seeks to infer users' private-attribute information according to their items list and recommendations. The recommender aims to extract users' interests while employing the attacker to regularize the recommendation process. Experiments show that the proposed model both preserves the quality of recommendation service and protects users against private-attribute inference attacks.

\end{abstract}

%
%

\ccsdesc[500]{Information systems~Recommender systems}
\ccsdesc[500]{Security and privacy~Social network security and privacy}
\ccsdesc[500]{Security and privacy~Privacy protections}

\keywords{Privacy-Aware Recommendation; Private-Attribute Protection; Adversarial Learning; Privacy; Utility}

\maketitle

\section{Introduction}
Recommendation systems play an important role in helping users find relevant and reliable information that is of potential interest~\cite{koren2009collaborative}. These systems build profiles that represent user's interests~\cite{konstan2012recommender,beigi2018similar} and recommend relevant items to the users based on the constructed profiles~\cite{rashid2002getting}. Despite the effectiveness of recommendation systems, they can be sources of user privacy breach. Existing work has shown that if malicious attackers have access to the system's output and unrestricted auxiliary information about their targets, they are able to extract their entire user-item interactions history~\cite{ramakrishnan2001privacy,calandrino2011you,mcsherry2009differentially,beigi2018privacy}. One main reason is that recommendation systems' outputs (i.e., product recommendation) are partially derived from other users' choices (i.e., user-item interactions history). Thus, privacy concerns arise. 

One of privacy issues is the re-identification attack where a malicious adversary attempts to infer user's actual ratings by seeking if a target user is in the database~\cite{beigi2018privacy}. Prior research on privacy preserving recommendation systems has extensively addressed this type of privacy breach. Common techniques include (1) modifying the output of the recommendation system algorithm so that the absence or presence of a single rating or an entire user data is masked (i.e., differential privacy based techniques)~\cite{mcsherry2009differentially,hua2015differentially,zhu2016differential}; and (2) coarsening the user's interactions history by adding dummy items and ratings such that the adversary cannot deduce the user's actual ratings and preferences (i.e., perturbation based techniques)~\cite{rebollo2011information,polat2003privacy,luo2014privacy}.

Another privacy issue is the disclosure of user private-attribute information through leaked users' interactions history~\cite{weinsberg2012blurme,beigi2019privacy}. Private attribute information contains those attributes that users do not wish to disclose such as age, gender, occupation and location. This type of privacy breach is known as the private-attribute inference attack in which the adversary's goal is to infer private attributes of target users given their interactions history. Little has been done to protect users against this attack of private-attribute inference~\cite{jia2018attriguard,weinsberg2012blurme,beigi2019not,beigi2019privacy} with focus on anonymizing user-item data before publishing it. Data obfuscation comes at the cost of utility loss where utility is defined as the quality of service users receive. The existing work addresses the utility loss by minimizing the amount of changes made to the data~\cite{jia2018attriguard,weinsberg2012blurme}. However, in the context of recommendation, the utility loss due to this approach can lead to degraded recommendation results. 
 Moreover, just sharing perfectly obfuscated user-item data with a recommendation system does not necessarily prevent the adversary from inferring users' private information in future when they receive and accept new recommendations (e.g., when purchasing new products).

 This research aims to devise a mechanism to counter private-attribute inference attacks in the context of recommendation systems. We propose a privacy-aware \textit{R}ecommender with \textit{A}ttribute \textit{P}rotection, namely \textsc{RAP}, which offers relevant products in a way that makes any inference of user's private attributes difficult from his interactions history and recommendations. 
 The proposed model seeks to concurrently prevent the leakage of users' private attribute information while retaining high utility for users. 
 
Recommendation while countering private-attribute inference attack can be naturally formulated as a problem of adversarial learning~\cite{goodfellow2014generative}. In our proposed \textsc{RAP}, there are two components: a Bayesian personalized ranking recommender and a private-attribute inference attacker (illustrated in Figure.~\ref{overview}). The private-attribute inference attacker seeks to accurately infer users' private attribute information. The attacker aims to iteratively adapt its model with respect to the existing recommender. 
The recommender extracts latent representations of users and items for personalized recommendation, and simultaneously utilizes the private-attribute inference attacker to regularize the recommendation process by incorporating necessary constraints to fool the attacker. Therefore, \textsc{RAP} optimizes a composition of two conflicting objectives, modeled as a min-max game between recommender and attacker components. Its objective is to recommend relevant, ranked items to users such that a potential adversary cannot infer their private attribute information.
 
In essence, we investigate the following research issues: (1) whether we can develop a personalized privacy-aware recommendation system to guard against private-attribute inference attacks; and (2) how we can ensure that the user's private attributes are effectively obscured after receiving personalized recommendation. 
Our research on these issues results in a novel framework \textsc{RAP} with the following main contributions:
\begin{itemize}[leftmargin=*]
    \item To the best of our knowledge, this is the first effort in proposing a recommendation system with guarding against the inference of private attribute information while maintaining the user utility.
    \item The proposed \textsc{RAP} model uses an attacker component that regularizes the recommendation process to protect users against private-attribute inference attack. 
	\item The proposed \textsc{RAP} model is a general framework for recommendation systems. Both of the integrated Bayesian personalized recommender and the private-attribute attacker can be easily replaced by different models designed for specific tasks.
	
	\item We conduct experiments on real-world data to demonstrate the effectiveness of \textsc{RAP}. 
	 Our empirical results show that \textsc{RAP} preserves user utility and privacy. The results demonstrates that \textsc{RAP} outperforms the state-of-the-art related work and enables an adjustable balance between private-attribute protection and personalized recommendation. 
\end{itemize}


\section{Problem Statement}

Before formally defining our problem, we first describe the notations used in this paper. Let $\mathcal{I} = \{i_1,...,i_{M}\}$ denotes items, and $\mathcal{U}= \{u_1,..., u_{N}\}$ denotes users. Also, $\mathcal{I}_h$ represents the set of items rated by user $h$, and $\mathcal{R}_h$ is set of items recommended to $h$. $\mathcal{P} =\{p_1,...,p_T\}$ denotes a set of $T$ private attributes (e.g., age, gender). $\mathbf{R}$ represents user-item rating matrix. The goal is to recommend products to people that would be interesting for them. However, we want to protect people's privacy against a malicious adversary who attempts to infer their private attribute information according to the user's list of items information. Items list $\mathcal{S}_h$ for each user $h$ is union of his previously rated and newly recommended items, i.e., $\mathcal{S}_h = \{\mathcal{I}_h \cup \mathcal{R}_h\}$. 
 In particular, the malicious attacker has a framework which takes a target user's interactions and infers the user's private attribute:
\begin{theorem}
	We aim to learn a function $f$ that can recommend interesting and relevant products $\mathcal{R}_h$ to each user $u_h$ such that, 1) the adversary cannot infer the targeted user's private attribute information $\mathcal{P}$ from the user's list of items information, $\mathcal{S}_h = \{\mathcal{I}_h \cup \mathcal{R}_h \}$ and 2) the set of recommended items $\mathcal{R}_h$ is interesting for the user. The problem can be formally defined as: $\mathcal{R}_h = f(\mathcal{I}_h, \mathbf{R}, \mathcal{P})$
\end{theorem}
Note that, the goal is to protect users against a malicious adversary who has access to the users' items list, but not against the recommender which is trusted.
\section{Related Work}
\begin{figure}[t]
	\centering
	\includegraphics[width=0.95\linewidth]{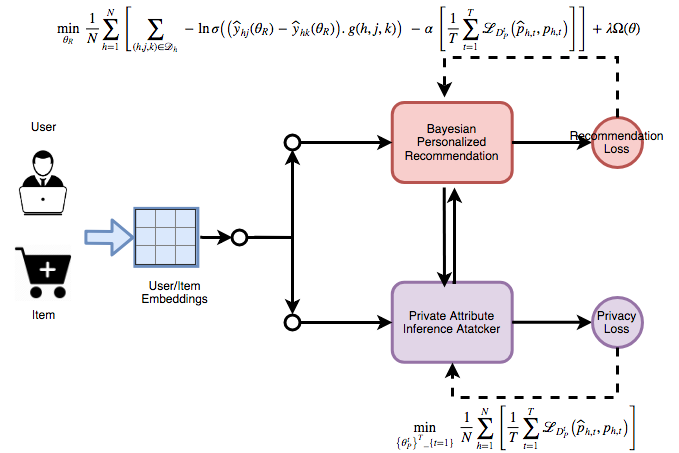}
	\caption{The architecture of Recommendation with Protection (\textsc{RAP}) with two components: a Bayesian personalized recommender and a private-attribute inference attacker.}\label{overview}
\end{figure}
Explosive growth of the Web has raised numerous challenges for online users including disinformation spread~\cite{alvari2018early,alvarihawkes,alvari2019less,alvariviolent} and threats to users' privacy~\cite{beigi2019identifying,beigi2018privacy}. Addressing user privacy issues has been studied from different aspects such as textual information~\cite{beigi2019privacy,beigi2019not}, web browsing histories~\cite{beigi2019protecting}, private-attributes disclosure~\cite{beigi2019privacy,jia2018attriguard} and recommendation systems~\cite{zhu2016differential,luo2014privacy} (for a comprehensive survey refer to~\cite{beigi2018privacy}). Our work
is related to a number of research which we discuss below while we elaborate on the differences between our work and them.

\noindent\textbf{Privacy and Recommendation Systems.}
Existing privacy preserving works in recommendation systems focus on protecting users against re-identification attacks in which an adversary tries to infer a targeted user's actual ratings and investigate if the target is in the database. They could be categorized into differential privacy based~\cite{mcsherry2009differentially,jorgensen2014privacy,hua2015differentially,zhu2016differential} and perturbation based~\cite{rebollo2011information,polat2003privacy,luo2014privacy} approaches. Some methods utilize differential privacy strategy~\cite{dwork2008differential} to modify the answers of the recommendation algorithm so the the presence of a user's data (either a single user-item rating or entire user's history) is masked by increasing the chance that two arbitrary records have close probabilities to generate the same noisy data.
McSherry et al.~\cite{mcsherry2009differentially} utilize differential privacy to construct private covariance matrices to be further used by recommender. Another work~\cite{jorgensen2014privacy} clusters users w.r.t. the social relations and generates differentially private average of users' preferences in each cluster. Hua et al.~\cite{hua2015differentially} propose a private matrix factorization which adds noise to item vectors to make them differentially private. Bassily et al.~\cite{bassily2015local} modify user-item ratings data to satisfy differential privacy and then share it with recommender. Another work~\cite{zhu2016differential} makes items list differentially private and then sends it to recommender. Perturbation based techniques obfuscate user's interactions history by adding fake items and ratings to it. 
Rebollo et al.~\cite{rebollo2011information} propose an information theoretic based privacy metric and then find the obfuscation rate for generating forged user profiles so that the privacy risk is minimized.
Similarly, \cite{parra2014optimal} proposes to 
 add or remove items and ratings from user profiles minimize privacy risk. Polat et al.~\cite{polat2003privacy} use a randomized perturbation technique by sharing disguised z-score for items a given user have rated. In another work~\cite{luo2014privacy}, similar users are grouped to each other. Aggregated ratings of the users within the same group is then used to estimate a group preference vector. Similar to~\cite{polat2003privacy}, randomness is then added to the preference vector to be shared with the recommender.

\noindent\textbf{Attribute Inference Attacks and Defenses}
Private-attribute inference attack focuses on inferring users' private attribute information from their publicly available information. These attacks could be categorized into three groups.
 A group of these attacks leverages a target user's friends' information~\cite{he2006inferring,lindamood2009inferring,gong2014joint} and community membership information~\cite{zheleva2009join,mislove2010you} to infer target's private attributes. 
 Second group of these attacks are those works which leverage users' behavioral information such as movie-rating behavior~\cite{weinsberg2012blurme} and Facebook likes~\cite{kosinski2013private} to infer their private attribute information. 
 The third group of works exploits both friend and behavioral information~\cite{gong2016you,gong2018attribute,jia2017attriinfer}. Gong et al.~\cite{gong2016you,gong2018attribute} make a social-behavior-attribute network in which all users' behavioral and friendship information is integrated in a unified framework. Private attributes are then inferred through a vote distribution attack model. Another work~\cite{jia2017attriinfer} incorporates structural and behavioral information from users who do not have the attribute in the training process, i.e. negative training samples.

Little work focuses on protecting users against private-attribute inference attacks~\cite{weinsberg2012blurme,jia2018attriguard}. In~\cite{weinsberg2012blurme}, a predefined number of dummy items is added to each user's profile which are negatively correlated with his actual attributes before publishing anonymized user-item ratings data. In a recent paper~\cite{jia2018attriguard}, after a value is sampled for the given private attribute w.r.t. a certain probability distribution which is different from the user's actual attribute, the minimum noise is found and added to the user-item data via adapting evasion attacks such that the malicious attacker predicts the sampled attribute value as the user's private attributes.

Our work is different from the existing works. First, existing privacy preserving recommendation systems do not specifically target the private-attribute inference attacks. Second, existing defenses against this attack~\cite{weinsberg2012blurme,jia2018attriguard} address the utility loss 
 by minimizing the amount of changes made to the data. However, in scope of recommendation systems, this approach can mean neglecting the quality of received services, i.e., poor recommendation results. Third, sharing anonymized data with recommender does not preclude the malicious attacker to infer private attribute information when users receive new recommendations. All of these limitations arises the need for having a recommendation system guarding against the inference of private attribute while maintaining the user utility.

\section{Recommendation with Attribute Protection (RAP)} 
Our proposed recommendation framework, \textsc{RAP}, aims to concurrently recommend interesting items to users and protect them against private attribute leakage. 
 The entire model is illustrated in Figure.~\ref{overview}. This framework consists of two major components, 1) a Bayesian personalized recommender, and 2) a private-attribute inference attacker. The personalized ranking recommender $D_R$ aims to extract users' actual preferences and recommend relevant items to them. The private-attribute inference attacker $D_P$ seeks to develop a model which can deduce users' private information w.r.t. the existing recommendation system. Recommendation component then utilizes $D_P$ to guide the recommendation process by ensuring that the union of previously rated and newly recommended items does not leak user's attributes and further fools the adversary in $D_P$. Inspired by adversarial machine learning, we model this objective as a min-max game between two components, i.e. attacker $D_P$ seeks to maximize its gain and recommender $D_R$ aims to minimize both its recommendation loss and attacker $D_P$'s gain. 
 The final output of \textsc{RAP} for each user, is a list of top-$K$ items which are interesting yet safe for them.

\subsection{Bayesian Personalized Recommendation}\label{sec:recommendation}
\begin{figure}[t]
	\centering
	\includegraphics[width=1\linewidth]{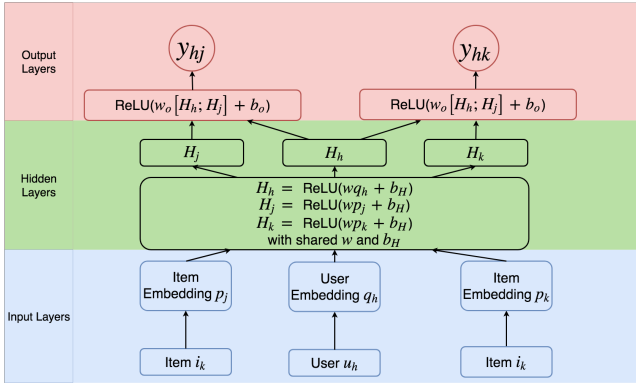}
	\caption{Overview of the Bayesian personalized recommendation component.}\label{bpr}
\end{figure}

In this section, we propose a new Bayesian personalized recommendation model. The proposed model structure is shown in Fig.~\ref{bpr}. This model first extracts users and items latent embeddings and then utilizes learning to rank approach to recommend items to users. 

Learning to rank methods have been introduced to optimize recommendation systems toward personalized ranking. Inspired by recent success of Bayesian Personalized Ranking (BPR)~\cite{rendle2009bpr} in image and friend recommendation systems~\cite{niu2018neural,ding2017baydnn}, we choose BPR aver other approaches. The idea behind BPR is that observed user-item interactions should be ranked higher than unobserved ones. Learning from implicit feedback, BPR goal is to maximize the margin between an observed user-item interaction and its unobserved counterparts. In particular, BPR behavior could be interpreted as a classifier in which given a positive triplet instance of user $h$ and items $j$ and $k$, $(h,j,k)$, it determines whether the user-item interaction $(h,j)$ should have a higher rank score than $(h,k)$.

This recomemndation component has three inputs, the user $h$ and items $j$ and $k$.
We denote the user and items indices by a tuple of vectors $(u_h, i_j, i_k)$ which are one-hot encodings of users and items. Since there are $N$ users and $M$ items, the dimensions of $u_h$, $i_j$, and $i_k$ are $M$, $N$ and $N$, respectively. Following the input layer, each input layer is fully connected to the corresponding embedding layer to learn the latent representation of the users and items, $\mathbf{q}_h \in \mathbb{R}^d$, $\mathbf{p}_j \in \mathbb{R}^d$, where $d$ is the number of dimensions. The embedding dimensions for both users and items are the same:
\begin{equation}
\mathbf{q}_h = \mathbf{W}_h u_h, ~~~~~~~~\mathbf{p}_j = \mathbf{W}_j i_j,~~~~~~~~~~\mathbf{p}_k = \mathbf{W}_k i_k
\end{equation}
where $\mathbf{W}_h$, $\mathbf{W}_j$ and $\mathbf{W}_k$ are embedding matrices for users and items. In the next layer, user and item embedding vectors are passed to the hidden layers $H_h$, $H_j$, and $H_k$ for further calculations. For example, the hidden layer produces $H_h$ for user $h$ as:
\begin{equation}
H_h = ReLU(w {q}_h+b_H)
\end{equation}
where $ReLU$ is simply defined as $ReLU(x) = \max(0,x)$ and $w$ and $b_H$ are the weights and bias for units, respectively.

Using $H_h$, $H_j$, and $H_k$, the next layer produces the user's preference $\hat{y}_{hj}$, $\hat{y}_{hk}$ toward items $j$ and $k$, respectively. For example:
\begin{equation}
\hat{y}_{hj} = ReLU(w_o[H_{h};H_{j}]+b_o)
\end{equation}
where $b_o$ is the bias parameter in the output layer. The activation function is $RelU$ function and $[.;.]$ represents concatenation. 
 Note that due to the model simplicity, all users share the same latent representation learning parameters $\{w, b_H\}$ and $\{w_o, b_o\}$ in the hidden layer and output layer, respectively.

We use BPR to learn how to rank in the problem of recommendation. The final objective function is to minimize the following loss function w.r.t. $\theta_R$:
\begin{equation}
\mathcal{L}_{D_R} = \frac{1}{N}\sum_{h = 1} ^N \sum_{(h, j, k) \in \mathcal{D}_h} \! \! \! \! \! \! \! \!- \ln \delta((\hat{y}_{hj}(\theta_R) - \hat{y}_{hk}(\theta_R)).g(h,j,k) )+\lambda_{\theta_R} \lVert \theta_R \rVert^2
\label{finalBPR}
\end{equation}
where, $g(h,j,k)$ is the ground truth value for our model training:
 \begin{equation}
     g(h,j,k) = \begin{cases}
      1, & \text{if user $u_h$ prefers item $i_j$ over item $i_k$} \\
      -1, & \text{otherwise}
    \end{cases}
 \end{equation}
where set $\mathcal{D}_h = \{ (h,j,k) | j \in \mathcal{I}_h \text{ and } k \in \mathcal{I}/\ \mathcal{I}_h\}$ also denotes the training pairwise instances in which $\mathcal{I}$ and $\mathcal{I}_h$ represent the whole set of items and the set of items rated by user $u$, respectively. Moreover, $y_{hj}$ is the actual rating that user $h$ gives to item $j$. $\theta_R$ is also defined as $\theta_R = \{\mathcal{W}_{\mathcal{U}}, \mathcal{W}_{\mathcal{I}}, w, b_H,w_o,b_o\}$ such that $\mathcal{W}_{\mathcal{U}} = \{\mathbf{W}_1,.., \mathbf{W}_N\}$ and  $\mathcal{W}_{\mathcal{I}} = \{\mathbf{W}_1,.., \mathbf{W}_M\}$ represent the set of embedding matrices for $N$ users and $M$ items, respectively. The proposed model considers the recommendation problem as a binary classification problem to ensure that the pairwise preference relations hold. 

After training the recommendation model, given a user $h$, for every item $j$ that the user has not rated, i.e., $j \in \{\mathcal{I}/\ \mathcal{I}_h\}$, his preference score $\hat{y}_{hj}$ is predicted by the recommender. In order to calculate the preference score $\hat{y}_{hj}$, we pass the tuple $(h,j,j)$ to the recommender, and get $\hat{y}_{hj}$ and $\hat{y'}_{hj}$ as the model's output. The final preference score of user $h$ toward item $j$ is calculated as $\hat{y}_{hj} = 0.5 (\hat{y}_{hj}+\hat{y'}_{hj})$. All of the unrated items will be then sorted based on their preference scores descendingly and the top-$K$ items are then returned as the recommendation $\mathcal{R}_h$ to the user.

\subsection{Training an Attacker against Inferring Private Attribute Information}
\begin{figure}[t]
	\centering
	\includegraphics[width=1\linewidth]{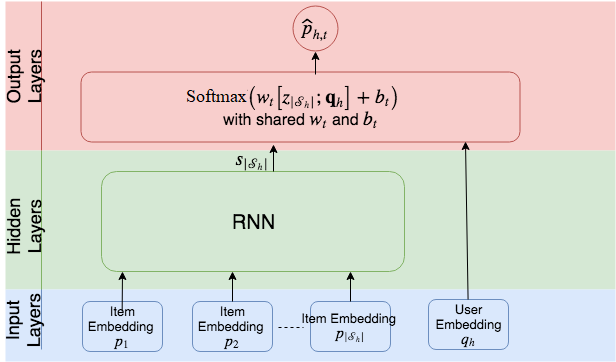}
	\caption{Overview of the private-attribute inference attacker component for one attribute.}\label{attack}
\end{figure}
The goal of our model is to recommend ranked items to users such that any potential adversary cannot infer users' private attribute information such as age, gender and occupation. However, a challenge is that the recommendation system does not know the malicious attacker's model. To address this challenge, we add a private-attribute inference attacker $D_P$ component to our model which seeks to learn a classifier that can accurately identify the private information of users from their previous interactions. Then, we leverage this component to regularize the recommendation process by incorporating necessary constraints in order to fool the adversary $D_P$ and further avoid the leakage of private attributes after recommendation. This part is discussed in details in Section.~\ref{total}.

The goal of the private-attribute attacker is now to predict target user $h$'s private attribute information by leveraging the information of his latent representation as well as the the latent representation of his items list. The user $h$'s items list $\mathcal{S}_h = \{\mathcal{I}_h \cup \mathcal{R}_h\}$ includes both items $\mathcal{I}_h$ that user has rated previously and new recommended items $\mathcal{R}_h$.
 Given $T$ private attributes (e.g., age, gender), the set of $\{\theta_{P^t}\}^T_{t=1}$ represents all the parameters included in the private-attribute inference attacker component $D_P$. The output of the private attribute attacker component for user $h$ w.r.t. $t$-th private attribute is the probability that user $h$ has $t$-th attribute.

We use $p_{h,t}$ to represent the actual value for user $h$'s $t$-th private attribute. The structure of private attribute inference attacker is represented in Fig.\ref{attack}.
The input to this model for each user $h$ is the latent embedding representations of each item $p_j$ in his items list $\mathbf{p}_j \in \mathcal{S}_h$, $j = 1,2 , ..., |\mathcal{S}_h|$ and $h$'s latent embedding representation $\mathbf{q}_h$. Given the input, the items embeddings are passed to a single-layer recurrent neural network (RNN) and the output of RNN ($z_{|\mathcal{S}_h|}$) is then concatenated with user's embedding. The last layer produces the predicted $t$-th sensitive attribute for user $h$, $\hat{p}_{h,t}$:
\begin{equation}
    \hat{p}_{h,t} = softmax(w_{t}[z_{|\mathcal{S}_h|};\mathbf{q}_h]+b_t)
\end{equation}
where $[.;.]$ represents concatenation. Also, $w_{t}$ and $b_t$ are the weights and bias for units, respectively and are shared among all users due to the model simplicity. We then minimize the private-attribute inference attacker component loss function $\mathcal{L}_{D_P^t}$ for all private attributes by seeking the optimal parameters $\{\theta_P^t\}^T_{t=1}$. The objective function for all users can be formally written as follows:
\begin{equation}
\mathcal{L}_{D_P} =  \frac{1}{N} \sum_{h=1}^N \bigg[\frac{1}{T} \sum_{t=1}^T \mathcal{L}_{D_P^t}(\hat{p}_{h,t}, p_{h,t})\bigg]\label{Priv}
\end{equation}
where $\mathcal{L}_{D_P^t}$ denotes the cross entropy loss 
for $t$-th private attribute.

\subsection{Adversarial Learning for Recommendation with Private-Attribute Protection}\label{total}
Thus far, we have discussed how we 1) learn users and items representations to recommend ranked items to each user based on his personalized preferences; and 2) train an attacker which can accurately infer a target user's private attribute information given a list of his rated items and received recommendations. 
 We stress that the adversary always has the upper hand and adapts his private-attribute inference attack in order to minimize his inference loss w.r.t. the existing recommendation system. The final objective is thus to recommend relevant ranked items to users such that a potential adversary cannot infer their private attribute information. To achieve two goals together, we design an optimization problem to minimize the recommendation loss of our model \textit{and} maximize the inference loss of a determined attacker who adaptively minimizes his loss. Inspired by the idea of adversarial learning, we model this optimization as a min-max game between two components, Bayesian personalized recommender and private-attribute attacker.
 
In our proposed model, the adversary tries to adapt itself and gets the maximum gain, while the recommendation system seeks to recommend ranked items to users. The recommended items not only align well with the users' preferences, but also minimize the adversary's gain. We reformulate the objective function of the recommendation system as minimizing attacker's gain and recommendation loss simultaneously:
\begin{equation}
	\underbrace{\min_{\theta_{R}}  \bigg (\mathcal{L}_{D_R} \ \overbrace{ - \alpha  \max_{\{\theta_P^t\}^T_{t=1}} \mathcal{L}_{D_P}}^\text{private-attribute attacker} \bigg)}_\text{privacy-aware recommendation system} \label{Obj_one}
\end{equation}

The inner part learns the most determined adversary which adaptively minimizes its loss regarding private-attribute inference given the users and items information. The outer part seeks to both minimize the recommendation loss and fool the given adversary. The parameter $\alpha$ controls the contribution of the private-attribute inference attacker in the learning process. Objective function in Eq.~\ref{Obj_one} can be written as follows:
\begin{align}
&\min_{\theta_{R}} \max_{\{\theta_P^t\}^T_{t=1}} \Bigg(\frac{1}{N} \sum_{h=1}^N \bigg[\sum_{(h, j, k) \in \mathcal{D}_h} - \ln \delta((\hat{y}_{hj}(\theta_R) - \hat{y}_{hk}(\theta_R)).g(h,j,k) ) \\\nonumber
&-\alpha  \bigg[\frac{1}{T} \sum_{t=1}^T \mathcal{L}_{D_P^t}(\hat{p}_{h,t}, p_{h,t})\bigg] \bigg]+ \lambda \Omega(\theta) \Bigg)\label{Obj_two}
\end{align}


\noindent where $\theta= \{\theta_{R}, \{\theta_P^t\}^T_{t=1}\}$ is the set of all parameters to be learned, $\Omega(\theta)$ is the $L_2$ norm regularizer on the parameters, and $\lambda$ is a scalar to control the contribution of the regularization $\Omega(\theta)$.




\subsection{Optimization Algorithm}
The optimization process is illustrated in Algorithm~\ref{alg:opt}. First, we create a mini-batch sample $\mathcal{U}_b$ of $b$ users from the training data and serve their private attribute and item-rating information to the model. Next, we train the Bayesian personalized recommender $D_R$ according to the Eq.~10 w.r.t. $\theta_R$ in Line~\ref{alg:rec}. 
 Then, for each user $h$ in $\mathcal{U}_b$ we calculate the top-$K$ recommended items $\mathcal{R}_h$ and accordingly make his list of items, $\mathcal{S}_h$. The private-attribute inference attacker component is then trained according to the users and item embeddings information using Eq.~\ref{Priv} in Line~\ref{alg:attack}. After training \textsc{RAP}, for each user $h$, a list of top-$K$ items $\mathcal{R}_h$ will be returned as recommendation.

\begin{algorithm}[t]
	\caption{The Learning Process of \textsc{RAP} model}\label{learning}
	\begin{algorithmic}[1]
		\REQUIRE~~ Items set $\mathcal{I}$, training user data $\mathcal{U}$, training user-item matrix data $\mathbf{R}$, batch size $b$, $\theta_R$, $\{\theta_P^t\}^T_{t=1}$, $\alpha$, $\lambda$ and $K$. 
        \ENSURE~~ Trained recommendation with protection \textsc{RAP}.
		\REPEAT \label{alg:repeat}
		\STATE Create a mini-batch $\mathcal{U}_b$ of $b$ users  with their private-attribute and item-rating information from $\mathcal{U}$
		\STATE Train the recommendation with attribute protection via Eq.~10 w.r.t. $\theta_R$ \label{alg:rec}
		\STATE For each user $h$ in $\mathcal{U}_b$, calculate the top-$K$ recommended items $\mathcal{R}_h$ \label{alg:item}
		\STATE Train the private-attribute inference attacker $D_P$ (i.e., $\{\theta_P^t\}^T_{t=1}$) via Eq.~\ref{Priv} given the users' information including their list of items information, i.e., $\mathcal{S}_h = \{\mathcal{I}_h \cup \mathcal{R}_h\}$ \label{alg:attack}
		\UNTIL{\text{Convergence}}  \label{alg:until}
	\end{algorithmic}\label{alg:opt}
\end{algorithm}
\section{Experiments}
In this section we conduct experiments to evaluate the efficiency of the proposed framework in terms of both privacy and quality of the recommendation. We aim to answer the following questions:
\begin{itemize} [leftmargin=*]
    \item \textbf{Q1} - \textit{Privacy}: How does \textsc{RAP} perform in preventing leakage of users' private information?
    \item \textbf{Q2} - \textit{Utility}: How does \textsc{RAP} perform in recommending relevant items to users? 
    \item \textbf{Q3} - \textit{Utility-Privacy Relation}: Does the improvement in privacy result in sacrificing the utility of recommendation system?
\end{itemize}

To answer the first question (\textbf{Q1}), we examine our model against different private information with different distributions, such as age, gender, and occupation. Then, we evaluate the effectiveness of \textsc{RAP} in preventing leakage of users' private information given union of users' previously rated and newly recommended items. Addressing leakage of private attribute information may result in recommendation performance deterioration. Therefore, to answer the second question (\textbf{Q2}), we examine the performance of \textsc{RAP} in terms of the quality of the recommendation. Finally, to answer the third question (\textbf{Q3}), we investigate the loss in recommendation performance when enhancing privacy of users. 

\subsection{Data}
We use publicly available data MovieLens~\cite{harper2016movielens}. 
This dataset includes $100,000$ ratings by 943 users on 1,682 movies. Each user has rated at least 20 movies and the rating scores are between 1 and 5. 
In the collected dataset, each user is associated with three private attributes, gender (male/female), age, and occupation. For this paper, we follow the setting of~\cite{hovy2015tagging} and categorize age attribute into three groups, over-45, under-35, and between 35 and 45. In total, 21 possible occupations have been considered for this data. The average number of rated items for each user is 129. 
\subsection{Experimental Setting}\label{exp-sett}
Here, we first explain how we design experiments to evaluate utility and privacy. Then, we discuss evaluation metrics and baselines.

\noindent \textbf{Implementation Details:} The parameters for recommendation and attacker components are determined through grid search. For the Bayesian personalized ranking recommendation component, we set the dimension of first layer as $d=70$. Accordingly, size of user and item embedding vectors is $d=70$. The dimension of hidden layer is also set as $20$. 
 For the private-attribute inference attacker component, we use single layer RNN with the dimension of input layer set as $d=70$. User and item embeddings are then passed from recommendation component to the attacker component. The dimension of hidden layer is set as $100$. The parameters $\alpha$ and $\lambda$ are also determined through cross-validation, $\alpha = 1$ and $\lambda=0.01$.

 We initialize the weight matrices in both components with random values uniformly distributed in $[0,1]$. The error gradient is back propagated from output to input and parameters in each layer are updated. The optimization algorithm used for gradient update is Adam's algorithm~\cite{kingma2014adam}. The loss generally converges after 20 epochs. The batch size we use in experiments is $b=32$. 

\noindent \textbf{Recommendation Evaluation:} We evaluate the performance of recommendation by examining the quality of recommended items for all users. We follow the setting of \cite{jia2018attriguard} to set-up the experimental settings. To do so, we split the data for train and test as follows. For each user $h$ in the data, we randomly select $l$ rated items for test set and the remaining $n_h - l$ items for training set, where $n_h$ is the number of rated items for user $h$. We set the item rating for those in the test set as zero. We vary the value of $l$ as $\{35,40,45\}$. Then, the top-$K$ items are then returned to each user as the recommendation. Note that we assume \textsc{RAP} has access to the users' private attribute information during the training process. 

\noindent \textbf{Private-Attribute Evaluation:} We evaluate privacy of users in terms of their robustness against the malicious attribute inference attacks in which the adversary's goal is to infer users' private attributes. In particular, the malicious attacker learns a multi-class classifier which takes a target user $h$ list of items information, i.e. $\mathcal{S}_h = \{\mathcal{I}_h \cup \mathcal{R}_h\}$, where $\mathcal{I}_h$ is set of $h$'s rated items and $\mathcal{R}_h$ is set of items recommended to $h$. The adversary then infers the user's private attributes, i.e., gender, age, and occupation. 

We use a Neural Network (NN) model as the adversary's classifier. Note that \textsc{RAP} is not aware of the adversary's model. In this attack, the adversary deploys a feed-forward network with a single hidden layer to perform the attack. The input to this model is one-hot encoding of each user, $\mathcal{S}_h = \{\mathcal{I}_h \cup \mathcal{R}_h\}$. Since there are $M$ items in the dataset, the dimension of input vector is $M$. The input layer is then fully connected to the hidden layer with dimension of hidden state set as $100$ and a $softmax$ layer used as the output layer. The dimension of the hidden layer is determined through grid search. We note that Gong et al.~\cite{nzhenqiang2016you} also proposed an attribute inference attack which leverages both social friends and rating behavior. However, their attack is not applicable to our problem as we focus on leveraging only user-item rating information.

We follow the setting of \cite{jia2018attriguard} to set-up the experiments. We split the data to train and test sets by sampling $80\%$ of the users in the dataset uniformly at random as the training set and use the remaining users as testing set. We assume that the users in the training set has publicly disclosed their private information while the users in the testing set keep those attribute information private. Then, for each user in the test set, we randomly select $l$ rated items and remove them from the user's rating history by setting the their rating as zero. We keep the user-item ratings for users in the training set intact (i.e., original user-item ratings). Trained \textsc{RAP} model is deployed on the users in the test set and top-$l$ recommended items $\mathcal{R}_h$ are added to the users' previously rated items $\mathcal{I}_h$, in order to make $\mathcal{S}_h = \{\mathcal{I}_h \cup \mathcal{R}_h\}$. We vary value of $l$ as $\{35,40,45\}$.

The adversary's classifier is trained on the training set and evaluated on the users in the test set. Note that we assume that the malicious attacker knows the original intact user-item interactions for those users in the training set and seeks to predict private attribute information of the users in the test set, given their $\mathcal{S}_h$. We evaluate a malicious attack for each private attribute.
\begin{table*}[ht]
\small
	\begin{tabular}{l p{0.65cm}p{0.65cm}p{0.65cm}p{0.65cm} p{0.65cm}|p{0.65cm}p{0.65cm}p{0.65cm} p{0.65cm}p{0.65cm}|p{0.65cm}p{0.65cm} p{0.65cm}p{0.65cm}p{0.65cm}}
		\toprule
		\multirow{2}{*}{Model} &
		\multicolumn{15}{c}{$\#$ test items ($l$)} \\
		& \multicolumn{5}{c}{35}  & \multicolumn{5}{c}{40} & \multicolumn{5}{c}{45} \\
		& Gen & Age & Occ & $P@K$ & $R@K$ & Gen & Age & Occ & $P@K$ & $R@K$ & Gen & Age & Occ & $P@K$ & $R@K$  \\
		\midrule
		
		\textbf{\textsc{Original}} &  0.7662 & 0.7050 & 0.8332 & 0.156 & 0.156 & 0.7662 & 0.7050 & 0.8332 & 0.151 & 0.172 & 0.7662 & 0.7050 & 0.8332 & 0.145 & 0.187  \\ 
		
		\textbf{\textsc{LDP-SH}} &  0.6587 & 0.6875 & 0.8076 & 0.071 & 0.071 & 0.6440 & 0.6777 & 0.7954 & 0.062 & 0.078 & 0.6398 & 0.6732 & 0.7817 & 0.055 & 0.081  \\
		\textbf{\textsc{BlurMe}} & 0.6266 & 0.6177 & 0.7614 & 0.118 & 0.118 & 0.6013 & 0.5949 & 0.7589 & 0.109 & 0.134 & 0.5884 & 0.5901 & 0.7522 & 0.099 & 0.150 \\
		\textbf{\textsc{RAP}} & \textbf{0.6039} & \textbf{0.5397} & \textbf{0.7319} & \textbf{0.152} & \textbf{0.152} & \textbf{0.5714} & \textbf{0.5270} & \textbf{0.7315} & \textbf{0.147} & \textbf{0.168} & \textbf{0.5278} & \textbf{0.5262} & \textbf{0.7312} & \textbf{0.142} & \textbf{0.183}  \\
		\bottomrule
	\end{tabular}
	\caption{\textbf{\textsc{RAP} Performance. Higher $P@K$ and $R@K$ values show higher utility, while lower AUC indicates higher privacy.}}\label{privacy_utility}
\end{table*}

\noindent \textbf{Evaluation Metrics:} We use the following metrics for evaluating \textsc{RAP} performance w.r.t. malicious private-attribute inference (i.e., privacy) and product recommendation (i.e., utility):
\begin{itemize}[leftmargin=*]
    \item \textbf{Private-Attribute Evaluation:} 
    Since distribution of data for different private attribute values is imbalance, we report micro-AUC~\cite{fawcett2006introduction} of the adversary's classifier. Micro-AUC~\cite{fawcett2006introduction} gives a more accurate assessment. Lower AUC demonstrates higher privacy in terms of obscuring private attributes.
    \item \textbf{Recommendation Evaluation:} We use standard metrics that are widely used in other related works~\cite{ziegler2005improving}, i.e., $P@K$ and $R@K$. 

\textbf{$P@K$}: $P@K$ represents the ratio of test cases which has been successfully recommended in a top-$K$ position in a ranking list to value of $K$. For each user, we measure $P@K$ as:
\begin{equation}
    P@K = \frac{|\{\text{test items}\} \cap \{\text{top-$K$ returned items}\}|}{K}
\end{equation}

\textbf{$R@K$}: $R@K$ defines the ratio of top-$K$ recommended items which are in the test set to the number of items to be recommended in the test. For each user in the data, we measure $R@K$ as follows:
\begin{equation}
    R@K = \frac{|\{\text{test items}\} \cap \{\text{top-$K$ returned items}\}|}{|\{\text{test items}\}|}
\end{equation}
We then report the average of $R@K$ and $P@K$ for all users in the dataset and set the number of returned items as $K=35$. 

 \end{itemize}

\noindent\textbf{Baseline Methods:} We compare \textsc{RAP} with the following baselines:
\begin{itemize}[leftmargin=*]
    \item \textbf{\textsc{Original}}: This baseline is a variant of \textsc{RAP} which recommends items for each user without incorporating the private-attribute inference attacker component, i.e., $\alpha = 0$.
    \item \textbf{\textsc{LDP-SH}~\cite{bassily2015local}}: This method adds noise to user-item ratings based on $\epsilon$-differential privacy. It requires categorical data which for our case, each user-item rating can be viewed as categorical data taking values $\{0,0.2, 0.4,0.6,0.8,1\}$.
    \item \textbf{\textsc{BlurMe}~\cite{weinsberg2012blurme}}: This method perturbs user-item ratings before sending to recommendation system. It adds new items to each user's profile that are negatively correlated with the user's actual private attributes and then adds the average rating score to those items. \textsc{BlurMe} needs to be deployed for each attribute separately.
\end{itemize}
To have a fair comparison between \textsc{RAP} and two baselines, we anonymze the user-item rating data w.r.t. baselines. The noisy manipulated data is used to train the recommendation model. We use matrix factorization model as the recommendation framework for both baselines. The discussed procedure is then used for evaluation.
\subsection{Privacy Analysis (Q1)}
The results against the malicious private-attribute inference attack (Section~\ref{exp-sett}) are demonstrated in Table.~\ref{privacy_utility}. We observe that increasing the number of test items ($l$) results in decrease of AUC score for all frameworks. This is because for each target user $h$ in the test set, $l$ recommended items $\mathcal{R}_h$ have been added to user's item list $\mathcal{S}_h$. Therefore, increase of $l$ can decrease the malicious attacker's chance for correctly inferring users' private attribute information. Moreover, \textsc{RAP} has significantly lower AUC score in comparison to \textsc{Original} for all three private attributes and thus outperforms \textsc{Original} in terms of obscuring users' private attribute information. \textsc{RAP} also has significantly better performance in hiding private information in comparison to \textsc{LDB-SH}. The reason is that \textsc{LDB-SH} aims to achieve a privacy goal that is different from preventing leakage of private information. This confirms that although adding noise and satisfying $\epsilon$-differential privacy can indirectly benefit private attribute leakage, it does not directly target this problem. These results show the importance of private-attribute inference attacker component in obfuscating private information. We also observe that \textsc{RAP} hides more private information rather than \textsc{BlurMe} (lower AUC score). This demonstrates that providing obfuscated user-item rating data to the recommendation system, does not necessarily guarantee preventing future private attribute leakage when user receives (and accordingly buy) more recommended products. Moreover, \textsc{BlurMe} needs to be deployed for each private attribute separately while \textsc{RAP} considers three private attributes all together.

These results confirm the efficiency of \textsc{RAP} in obscuring users' private attribute information and demonstrate that despite the fact that \textsc{RAP} is not aware of the adversary's inference model, it is prepared against the malicious attacker.




\begin{table*}[ht]
\small
	\begin{tabular}{l p{0.65cm}p{0.65cm}p{0.65cm}p{0.65cm} p{0.65cm}|p{0.65cm}p{0.65cm}p{0.65cm} p{0.65cm}p{0.65cm}|p{0.65cm}p{0.65cm} p{0.65cm}p{0.65cm}p{0.65cm}}
		\toprule
		\multirow{2}{*}{Model} &
		\multicolumn{15}{c}{$\#$ test items ($l$)} \\
		& \multicolumn{5}{c}{35}  & \multicolumn{5}{c}{40} & \multicolumn{5}{c}{45} \\
		& Gen & Age & Occ & $P@K$ & $R@K$ & Gen & Age & Occ & $P@K$ & $R@K$ & Gen & Age & Occ & $P@K$ & $R@K$  \\
		\midrule
		
		\textbf{\textsc{RAP}} &  0.6039 & \textbf{0.5397} & \textbf{0.7319} & \textbf{0.152} & \textbf{0.152} & 0.5714 & \textbf{0.5270} & \textbf{0.7315} & \textbf{0.147} & \textbf{0.168} & 0.5278 & \textbf{0.5262} & \textbf{0.7312} & \textbf{0.142} & \textbf{0.183}  \\ \hline
		
		\textbf{\textsc{RAPAge}} &  0.6450 & 0.5948 & 0.7528 & 0.150 & 0.150 & 0.5489 & 0.5938 & 0.7522 & 0.146 & 0.167 & 0.5475 & 0.5909 & 0.7497 & 0.141 & 0.182  \\
		\textbf{\textsc{RAPGen}} & \textbf{0.5332} & 0.6789 & 0.7558 & 0.151 & 0.151 & \textbf{0.5298} & 0.6614 & 0.7556 & 0.145 & 0.166 & \textbf{0.5211} & 0.6415 & 0.7555 & 0.141 & 0.181 \\
		\textbf{\textsc{RAPOcc}} & 0.6571 & 0.6949 & 0.7468 & 0.147 & 0.147 & 0.6485 & 0.6871 & 0.7466 & 0.141 & 0.161 & 0.6454 & 0.6853 & 0.7438 & 0.135 & 0.174 \\
		\bottomrule
	\end{tabular}
	\caption{\textbf{Impact of different private-attribute attacker components on \textsc{RAP} in terms of utility and privacy.}}\label{components}
\end{table*}

\begin{figure*}[t]
	\centering
	\small
		\subfloat[Attribute Age]{\includegraphics[scale=0.125]{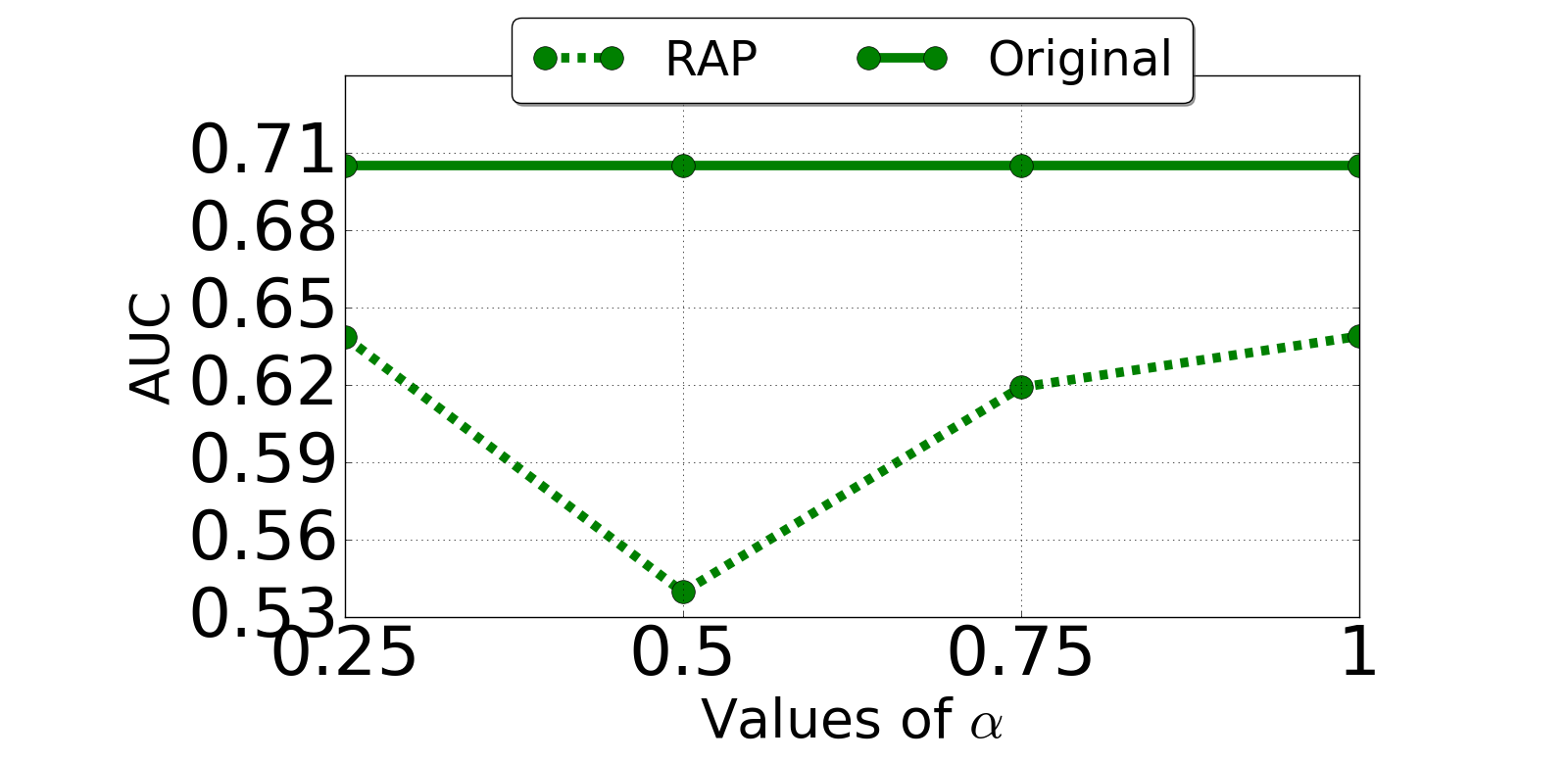}}\hspace*{-2em}
		\subfloat[Attribute Gender]{\includegraphics[scale=0.125]{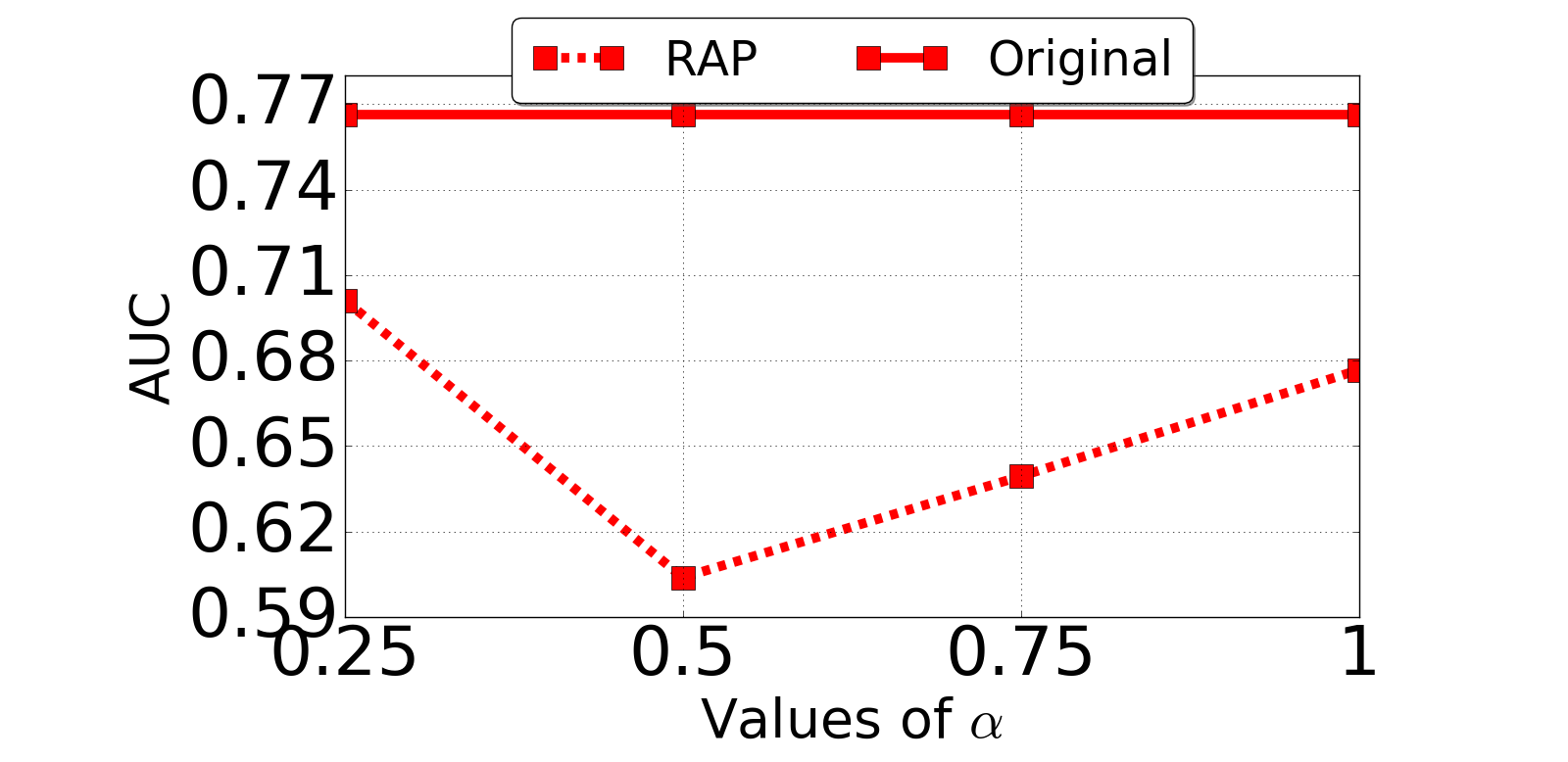}}\hspace*{-2em}
        \subfloat[Attribute Occupation]{\includegraphics[scale=0.125]{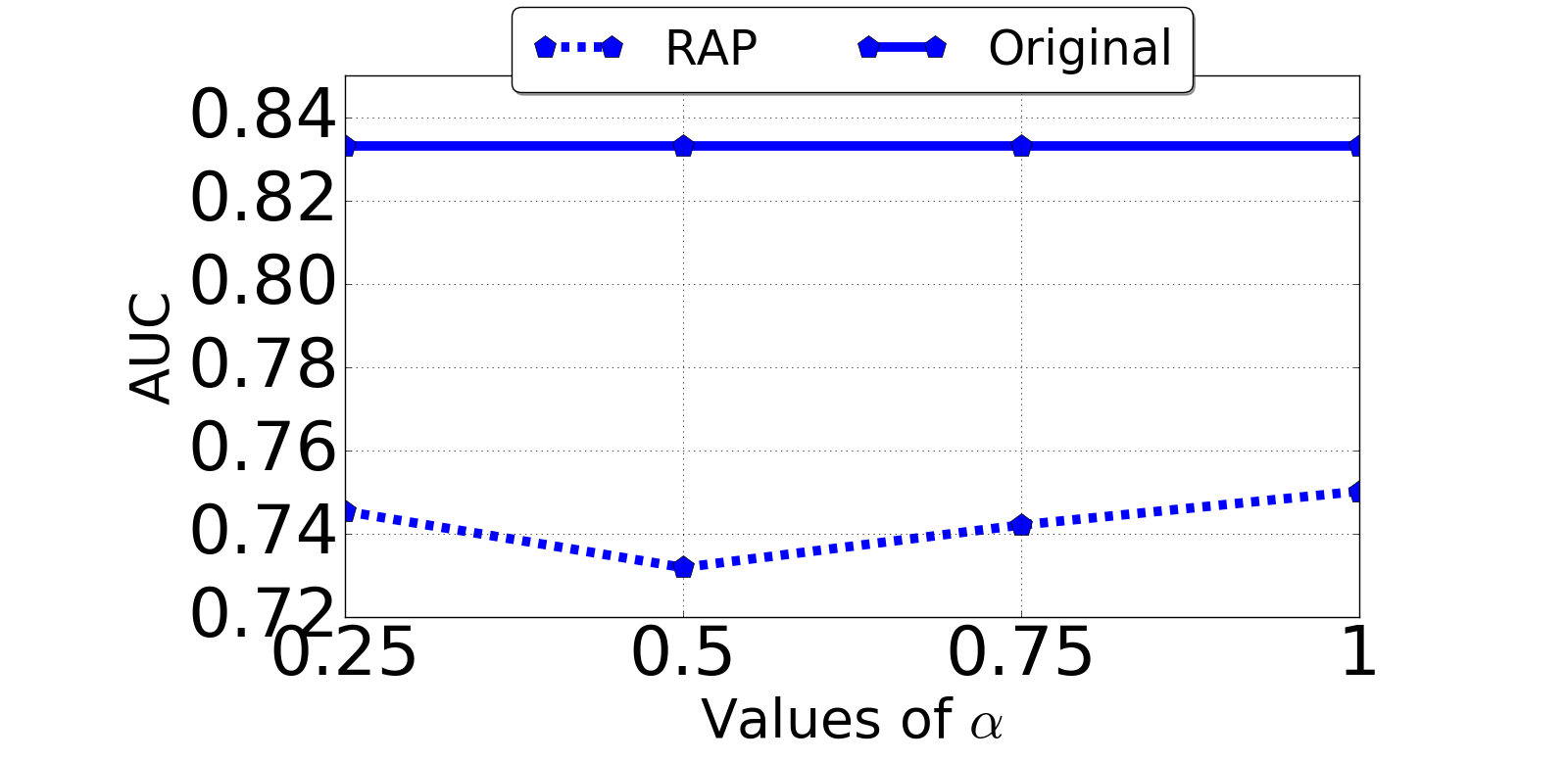}}\hspace*{-2em}
        \subfloat[Recommendation]{\includegraphics[scale=0.125]{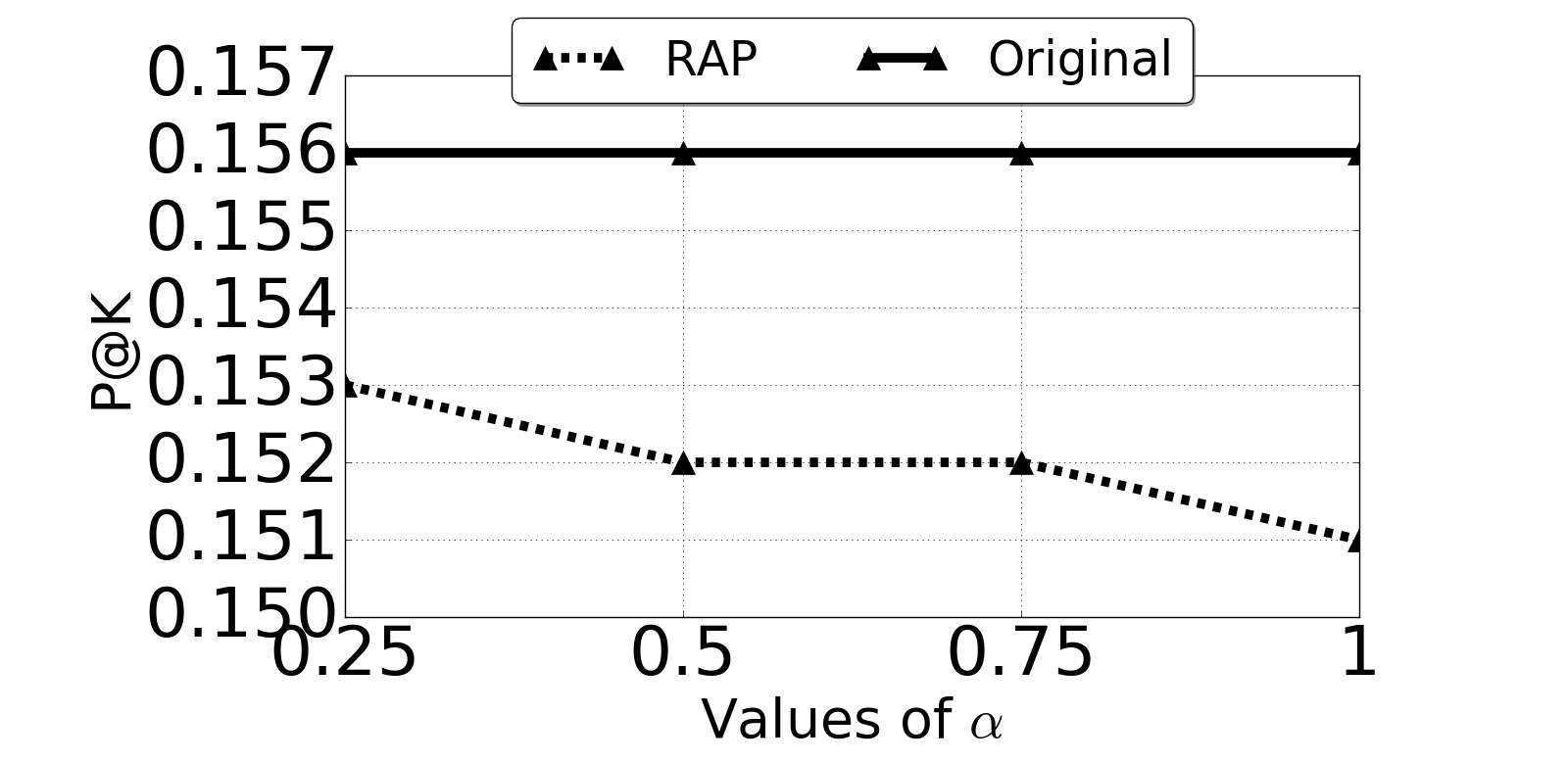}}
		\caption{\textbf{Performance results for private-attribute inference attack and recommendation task for different values of $\alpha$}}\label{para}
\end{figure*}
\subsection{Utility Analysis (Q2)}
The results for recommendation task for different methods and different number of test items ($l$) are shown in Table.~\ref{privacy_utility}. We observe that increasing the number of test items ($l$) results in increasing $R@K$ and decreasing $P@K$ for all methods. Note that the higher the $P@K$ and $R@K$ score values are, the higher recommendation quality is. 
 Another observation is that \textsc{LDP-SH} has the worst performance amongst all methods, i.e., lowest $P@K$ and $R@K$ scores. This is because of the way \textsc{LDP-SH} adds noise to the user data without considering the quality of recommendation service in practice which can result in degraded recommendation results. \textsc{BlurMe} has also lower performance than \textsc{RAP} as it neglects quality of recommendation results. These results confirm the effectiveness of Bayesian personalized recommendation component which helps \textsc{RAP} to take the utility into consideration in practice. Moreover, quality of recommendation results for \textsc{RAP} method is comparable to the \textsc{Original} approach. This means that \textsc{RAP} can accurately capture users' actual preferences and interests (i.e., high utility).

The results confirm the effectiveness of \textsc{RAP} in understanding users' actual preferences and recommending ranked relevant products that are interesting yet safe products to users.



\subsection{Utility-Privacy Relation (Q3)}
We compare the privacy and utility results in Table.~\ref{privacy_utility} for all methods. We observe that \textsc{LDP-SH} has the worst results in terms of both preserving privacy and recommendation performance. 
 Another observation is that \textsc{BlurMe} improves privacy compared to the \textsc{Original} method, but it loses utility in terms of recommendation system performance. This is in contrast with the results of \textsc{RAP}, which has outperformed \textsc{BlurMe} and \textsc{LDP-SH} in terms of recommendation and has comparable results with \textsc{Original}. \textsc{RAP} has also achieved the lowest AUC score and therefore highest privacy among all other methods. Comparing \textsc{RAP} with other methods confirms that approaching utility loss by minimizing the amount of data changes results in loss of quality of recommendation system in practice. 
This is reflected as degraded recommendation results for baseline approaches. Moreover, these results confirm the effectiveness of Bayesian personalized recommendation component in \textsc{RAP}, which helps us to consider quality of recommendation in practice. Results also demonstrate the complementary roles of both recommendation and private attribute components which guide each other through both privacy and utility issues. This results in a privacy-aware recommendation system which is prepared for private attribute inference attack and understands users' preferences.

\subsection{Impact of Different Components}
Here, we investigate the impact of different private attribute components on obscuring users' private information. We define three variants of our proposed framework, i.e., \textsc{RAPAge}, \textsc{RAPGen}, and \textsc{RAPOcc}. In each of these variants, the model is trained with the corresponding private-attribute inference attacker component, e.g. \textsc{RAPAge} is trained solely with age inference attacker component and does not utilize any other private-attribute attackers during training phase. Results are shown in Table.~\ref{components}. We observe that for gender attribute, \textsc{RAPGen} has the best performance in terms of obscuring gender attribute comparing to the other approaches (i.e., lowest AUC score). This is in contrast to quality of \textsc{RAPGen} performance for recommendation task which is lower than original proposed model \textsc{RAP}. For other private attributes, \textsc{RAP} still outperforms \textsc{RAPOcc} and \textsc{RAPAge} in terms of obscuring age and occupation attributes. Moreover, results show that using one private-attribute attacker compromises the effectiveness model for obfuscating other private attributes. 
 For the recommendation task, we surprisingly observe that using solely one of the private-attribute attackers in training process can result in performance reduction in comparison to \textsc{RAP} in terms of $P@K$ and $R@K$. This means that focusing merely on obscuring one private attribute can result in more recommendation performance degradation.

\subsection{Probing Further}
\textsc{RAP} has one important parameter $\alpha$ which controls the contribution from private-attribute attacker component. In this section, we probe further to investigate the effect of this parameter by varying it as $\{0.25, 0.5, 0.75, 1\}$. For this experiment, we set the number of test items $l=35$. We also set the number of top-$K$ returned items as $K=35$ for calculating $P@K$. Note that $P@K$ and $R@K$ are equal in this scenario as $K = l = 35$. Results are shown in the Fig.~\ref{para}.

Although $\alpha$ controls the contribution of private-attribute inference attacker component, we surprisingly observe that with the increase of $\alpha$, the AUC score for attribute inference attack decreases at first up to the point that $\alpha = 0.5$ and then it increases. This means that private information were obscured more accurately at the beginning with the increase of $\alpha$ and less later. Moreover, with the increase of $\alpha$, the performance of recommendation task decreases, i.e., lower $P@K$. This shows that increasing the contribution of private-attribute attacker component leads to decrease in the quality of recommendation framework. Another observation is that setting $\alpha = 0.25$ leads to improvement in hiding private attribute information in comparison to the results of using \textsc{Original} (or when $\alpha = 0$). This result shows the importance of the \textsc{RAP}'s private-attribute attacker component in preserving privacy of users. Another observation is that after $\alpha = 0.5$, continuously increasing $\alpha$ increases the AUC for malicious private-attribute inference attack, i.e., degrades the performance of hiding private information. The reason is that the model could overfit by increasing the value of $\alpha$ and lead to an inaccurate estimation of privacy protection.
\section{Conclusion}
In this paper, we propose an adversarial learning-based recommendation with attribute protection model, \textsc{RAP}, which guards users against private-attribute inference attack while maintaining utility. \textsc{RAP} recommends interesting yet safe products to users such that a malicious attacker cannot infer their private attribute from users' interactions history and recommendations. \textsc{RAP} has two main components, Bayesian personalized recommender, and private-attribute inference attacker. Our empirical results show the effectiveness of \textsc{RAP} in both protecting users against private-attribute inference attacks and preserving quality of recommendation results. \textsc{RAP} also consistently achieves better performance compared to the state-of-the-art related work. One extension to this work is to study the possibility of extending differential privacy mechanism for this type of attack in recommender systems. It would be also interesting to investigate personalized utility-privacy trade-off by tweaking framework parameters to fit the specific needs of individuals.
\begin{acks}
This material is based upon the work supported, in part, by NSF \#1614576, ARO W911NF-15-1-0328 and ONR N00014-17-1-2605.
\end{acks}

\bibliographystyle{ACM-Reference-Format}

\end{document}